\documentclass[conference]{IEEEtran}

\usepackage{amsmath,amsfonts,bm,amssymb,psfrag,ifthen,color,subfigure}
\usepackage{mathrsfs}

\usepackage[dvips]{epsfig}
\usepackage[dvips]{graphicx}
\usepackage{algpseudocode}
\usepackage{algorithm}
\usepackage{algorithmicx}
\usepackage{setspace}
\usepackage{cite}
\usepackage{url}
\usepackage{longtable}
\usepackage{stfloats}  
\usepackage{graphics,booktabs,color,subfigure}
\usepackage{float}

\begin{document}
\title{Robust Beamforming Design for Rate Splitting Multiple Access-Aided MISO Visible Light Communications}
\author{Shuai Ma$^\dag$,  Guanjie Zhang$^\dag$, Zhi Zhang$^\dag$, and Rongyan Gu$^\dag$ \\
$^\dag$ Sch. of Inform. and Cont.  Eng., China
University of Mining and Technology, Xuzhou 221116 \\

}

\maketitle
\begin{abstract}

In this paper, we focus on the optimal  beamformer design
for rate splitting multiple access (RSMA)-aided  multiple-input single-output
(MISO) visible light communication (VLC) networks.
First, we derive the closed-form lower bounds of the achievable rate of each user, which are the first theoretical bound of achievable rate for RSMA-aided VLC networks.
Second,  we investigate the optimal beamformer design for RSMA-aided VLC networks to maximize the sum rate under the optical and electrical power constraints.
In addition, we show that the proposed   RSMA-aided networks can achieve superior performance compared with space-division multiple access (SDMA) and non-orthogonal multiple access (NOMA).


%

\end{abstract}

\IEEEpeerreviewmaketitle
\section{Introduction}

 By utilizing  the deployed light
emitting diode (LED) as the transmitter,
   VLC can  provide
illumination and high-speed  communication simultaneously, without introducing interference to
radio-frequency (RF) communications\cite{Jovicic_ComMag_2013,Pathak_Surveys_2015
}.
Rate-splitting multiple
access (RSMA) was proposed \cite{Mao_EURASIP_2018}, which performs linearly precoded
rate splitting  at the transmitter and successive
interference cancellation (SIC)  at the receivers.  RSMA is a promising solution for  VLC networks. Two existing works have initiated the study of the RSMA design in VLC networks \cite{Tao_ICC_2020,Naser_Shimaa_2020}.
However, the achievable rate of RSMA-aided VLC networks is still unknown due to the distinct characteristics of VLC.
To be specific,  both peak and average optical power in VLC networks should be limited for
eye safety  and  practical illumination considerations \cite{Jovicic,Pathak}.

Motivated by the limitations of existing works on RSMA-aided VLC networks\cite{Tao_ICC_2020,Naser_Shimaa_2020,Mao_EURASIP_2018,Jovicic,Pathak},
  we focus on the  two key fundamental issues of  RSMA-aided VLC networks:  identifying achievable rates and developing both optimal and robust beamformer design.
   Specifically, by considering practical power constraints,  i.e., peak optical power constraint, average optical power constraint, and average electrical power constraint, we derive the lower bound of achievable rates of users in RSMA-aided VLC networks. Based on the derived lower bound of achievable rates, we investigate the optimal beamformer design for RSMA-aided VLC networks to maximize the sum rate under the optical and electrical power constraints of LEDs.Simulation results show that the proposed algorithms of RSMA-aided VLC networks can achieve superior performance compared with several baseline schemes.

\section{System Model of Downlink RSMA-aided VLC Network}


For the considered downlink RSMA-aided VLC network, the VLC base station (BS) equipped with $N$ LEDs
simultaneously serves $K$ single-access point  users by adopting RSMA scheme.
Specifically, at BS, the message ${\text{M}_k}$ intended to user-$k$  is split into a common message $\text{M}_{k,0}$ and a private message $\text{M}_k^p$, $\forall k \in {\cal{K}}$. Then, all the common messages  of $K$ users   $\left\{ {{{\rm{M}}_{k,0}}} \right\}_{k = 1}^K$ are combined into a
 super common message ${\text{M}_0}$, i.e., ${{\text M}_0}{\rm{ }} \buildrel \Delta \over = \left\{ {{{\rm{M}}_{1,0}}, \ldots ,{{\rm{M}}_{K,0}}} \right\}$.
  In addition, these $K+1$ signals $\left\{ {{s_i}} \right\}_{i = 0}^K$ satisfy $\left| {{s_i}} \right| \le {A_i}$, $\mathbb{E}\left\{ {{s_i}} \right\} = 0$ and $\mathbb{E}\left\{ {s_i^2} \right\} = {\varepsilon _i}$, $\forall i \in {\cal{I}}$, where $A_i > 0$ denotes  the signal amplitude. Therefore, the transmitted signal vector ${\bf{x}} = {\left[ {{x_1}, \ldots ,{x_N}} \right]^T}$ of VLC BS is given by
\begin{align}
{\bf{x}} = \sum\limits_{i = 0}^K {{{\bf{w}}_i}{s_i} + } {\bf{b}}.
\end{align}
The average electrical power and optical power of the transmitted signal ${\bf{x}}$ are respectively given by
\begin{subequations}
\begin{align}
&{P_{\rm{e}}} = {\mathbb{E}}\left\{ {{{\left\| {\bf{x}} \right\|}^2}} \right\} = \sum\limits_{i = 0}^K {{\varepsilon _i}{{\left\| {{{\bf{w}}_i}} \right\|}^2} + N{b^2}},\\
&P_{\rm{o}}^{{\rm{ave}}} = {\mathbb{E}}\left\{ {\bf{x}} \right\} = Nb.
\end{align}
\end{subequations}

Let $I_{\rm{H}}$ denotes the maximum permissible current of LEDs, the beamformer ${{\bf{w}}_i}$ should satisfy
 \begin{align}
\sum\limits_{i = 0}^K {{A_i}{\bf{w}}_i^T{{\bf{e}}_n} + b \le {I_{\rm{H}}},\forall n \in {\cal{N}}},
\end{align}
where ${\bf{e}}_n$ is an unit vector with the $n$th element equal to 1.

The optical channel between LED and user is dominated by line-of-sight (LOS) link, while diffuse links can be neglected \cite{T_Fath_2013,T_Q_Wang_2013}. Specifically, let ${{\bf{g}}_k} \buildrel \Delta \over = {\left[ {{g_{k,1}}, \ldots ,{g_{k,N}}} \right]^T}$ denote the channel gain vector of user-$k$, where ${g_{k,n}}$ is the channel gain between the $n$th LED and the user-$k$.
Therefore, at user-$k$, the received signal $y_k$ is given as
\begin{align} \label{received_signal}
{y_k} = \underbrace {{\bf{g}}_k^T{{\bf{w}}_0}{s_0}}_{{\rm{common}}\;{\rm{signal}}} &+ \underbrace {{\bf{g}}_k^T{{\bf{w}}_k}{s_k}}_{{\rm{private}}\;{\rm{signal}}} \nonumber\\
&+ \underbrace {\sum\limits_{i = 1,i \ne k}^K {{\bf{g}}_k^T{{\bf{w}}_i}{s_i}} }_{{\rm{interference}}} + {\bf{g}}_k^T{\bf{b}} + {z_k},
\end{align}
where ${z_k}$ is the received noise which follows the Gaussian distribution with mean zero and covariance $\sigma _k^2$. The term ${\bf{g}}_k^T{\bf{b}}$ denotes the DC component, which can be removed by the capacitor.

 Thus, the residual received signal of user-$k$ after SIC process can be represented
\begin{align}
y_k^{{\rm{SIC}}} = \underbrace {{\bf{g}}_k^T{{\bf{w}}_k}{s_k}}_{{\rm{desired}} \; {\rm{private}} \;{\rm{signal}}} + \underbrace {\sum\limits_{i = 1,i \ne k}^K {{\bf{g}}_k^T{{\bf{w}}_i}{s_i}} }_{{\rm{interference}}} + {\bf{g}}_k^T{\bf{b}} + {z_k}.
\end{align}

\section{DOWNLINK RSMA-aided VLC NETWORK}



 Let ${R_{k,c}}$ denotes the achievable  rate of decoding common signal ${{s_0}}$, and its lower bound is given by
\begin{align}
{R_{k,{\rm{c}}}}  \ge&\frac{1}{2}{\log _2}\left( {\frac{{2\pi \sigma _k^2 + \sum\limits_{i = 0}^K {{{\left| {{\bf{g}}_k^T{{\bf{w}}_i}} \right|}^2}{e^{1 + 2\left( {{\alpha _i} + {\gamma _i}{\varepsilon _i}} \right)}}} }}{{2\pi \sigma _k^2 + 2\pi \sum\limits_{j = 1}^K {{{\left| {{\bf{g}}_k^T{{\bf{w}}_j}} \right|}^2}{\varepsilon _j}} }}} \right).\label{common_rate}
\end{align}
 Let ${R_{k,{\rm{p}}}}$ denote the achievable   rate of decoding   private signal $s_k$,  and its lower bound is given by
\begin{align}
{R_{k,{\rm{p}}}} \ge &\frac{1}{2}{\log _2}\left( {\frac{{2\pi \sigma _k^2 + \sum\limits_{i = 1}^K {{{\left| {{\bf{g}}_k^T{{\bf{w}}_i}} \right|}^2}{e^{1 + 2\left( {{\alpha _i} + {\gamma _i}{\varepsilon _i}} \right)}}} }}{{2\pi \sigma _k^2 + 2\pi \sum\limits_{j = 1,j \ne k}^K {{{\left| {{\bf{g}}_k^T{{\bf{w}}_j}} \right|}^2}{\varepsilon _j}} }}} \right).\label{private_rate}
\end{align}

To ensure that ${{s_0}}$ is successfully decoded by all users, the transmission   rate of common message should not exceed ${R_{\rm{c}}} = \min \left\{ {{R_{1,{\rm{c}}}}, \ldots, {R_{K,{\rm{c}}}}} \right\}$.
Let
${c_k} \ge 0$
 denotes   one portion of  achievable  rate of common message  of user-$k$, where $\sum\nolimits_{k = 1}^K {{c_k} = {R_{\rm{c}}}}$ \cite{Mao_EURASIP_2018}.
Therefore, the achievable rate of user-$k$, denoted as $R_{k}$, can be expressed by
\begin{align}
R_{k}=c_{k} + {R_{k,{\rm{p}}}}, ~\forall k \in \cal{K}.
\end{align}

\subsection{Optimal Beamformer Design}
 Mathematically, the sum rate maximization problem of RSMA-aided VLC networks can be formulated as the following optimization problem
\begin{subequations}\label{p0}
\begin{align}
\mathop {\max }\limits_{\scriptstyle\left\{ {{{\bf{w}}_i}} \right\}_{i = 0}^K\hfill\atop
\scriptstyle\left\{ {{c_k}} \right\}_{k = 1}^K\hfill}& \sum\limits_{k = 1}^K {{R_k}}  \\
{\rm{s}}{\rm{.t.}}&\sum\limits_{k = 1}^K {{c_k} \le {R_{k,{\rm{c}}}}},~{c_k} \ge 0, \forall k \in \cal{K},\label{p0_c2}\\
&\sum\limits_{i = 0}^K {{{\left\| {{{\bf{w}}_i}} \right\|}^2}} {\varepsilon _i} \le {P_{\rm{t}}},\label{p0_c4}\\
&\sum\limits_{i = 0}^K {{A_i}} {\bf{w}}_i^T{{\bf{e}}_n} \le \min \left\{ {b,{I_{\rm{H}}} - b} \right\},\forall n \in {\cal{N}}.\label{p0_c5}
\end{align}
\end{subequations}

Since the objective function is non-convex, problem \eqref{p0} is computationally intractable.
To transform problem \eqref{p0} into tractable and equivalent problem, we
 first introduce a number of auxiliary variables   as follows
\begin{subequations}\label{introduce_variable}
\begin{align}
&{\widehat {\bf{w}}} \buildrel \Delta \over = {\left[ {{\bf{w}}_{_{0}}^{\rm{T}}, \cdots ,{\bf{w}}_{_{K}}^{\rm{T}}} \right]^{\rm{T}}},\\
&\widehat {\bf{d}} \buildrel \Delta \over = {\left[ {\varepsilon _{_0}^{1/2}, \cdots ,\varepsilon _{_K}^{1/2}} \right]^{\rm{T}}} \otimes {{\bf{1}}_N},\\
&{{\bf{a}}_n} \buildrel \Delta \over = {\left[ {{A_0}, \ldots {A_K}} \right]^T} \otimes {{\bf{e}}_n},\\
&{{\bf{G}}_{{\rm{c}},k}} \buildrel \Delta \over = {\rm{diag}}\left\{ {{\tau _0}, \cdots ,{\tau _K}} \right\} \otimes \left( {{{\bf{g}}_k}{\bf{g}}_k^T} \right),\\
&{\overline {\bf{G}} _{{\rm{c}},k}}{\rm{ }} \buildrel \Delta \over = 2\pi {\rm{diag}}\left\{ {0,{\varepsilon _1}, \cdots ,{\varepsilon _K}} \right\} \otimes \left( {{{\bf{g}}_k}{\bf{g}}_k^T} \right),\\
&{\widehat {\bf{G}}_{{\rm{c}},k}} \buildrel \Delta \over = {\rm{diag}}\left\{ {0,{\tau _1}, \cdots ,{\tau _K}} \right\} \otimes \left( {{{\bf{g}}_k}{\bf{g}}_k^T} \right),\\
&{\widehat {\bf{G}}_{{\rm{p}},k}} \buildrel \Delta \over = 2\pi {\rm{diag}}\left\{ {0,{\varepsilon _1}, \cdots {\varepsilon _{i - 1}},0,{\varepsilon _{i + 1}}, \cdots ,{\varepsilon _K}} \right\} \otimes \left( {{{\bf{g}}_k}{\bf{g}}_k^T} \right),
\end{align}
\end{subequations}
where ${{\bf{1}}_N}$ is a $N \times 1$ vector with all the elements equal $1$, and ${\tau _i} = {e^{1 + 2\left( {{\alpha _i} + {\gamma _i}{\varepsilon _i}} \right)}},\forall i \in {\cal{I}}$.

Based on the definitions \eqref{introduce_variable},  problem \eqref{p0} can be equivalently reformulated as follows
\begin{subequations}\label{p1}
\begin{align}
\mathop {\max }\limits_{\scriptstyle\left\{ {{t_k},{c_k}} \right\}_{k = 1}^K\hfill\atop
\scriptstyle\;\;\;\;{\bf{\hat w}}\hfill} &\sum\limits_{k = 1}^K {{t_k}} \\
{\rm{s}}.{\rm{t}}.&\frac{1}{2}{\log _2}\left( {2\pi \sigma _k^2 + {{\widehat {\bf{w}}}^T}{{\widehat {\bf{G}}}_{{\rm{c}},k}}\widehat {\bf{w}}} \right)\nonumber\\
&- \frac{1}{2}{\log _2}\left( {2\pi \sigma _k^2 + {{\widehat {\bf{w}}}^T}{{\widehat {\bf{G}}}_{{\rm{p}},k}}\widehat {\bf{w}}} \right) \ge {t_k} - {c_k},\label{p1_c1}\\
&\frac{1}{2}{\log _2}\left( {2\pi \sigma _k^2 + {{\widehat {\bf{w}}}^T}{{\bf{G}}_{{\rm{c}},k}}\widehat {\bf{w}}} \right)\nonumber\\
&- \frac{1}{2}{\log _2}\left( {2\pi \sigma _k^2 + {{\widehat {\bf{w}}}^T}{{\overline {\bf{G}} }_{{\rm{c}},k}}\widehat {\bf{w}}} \right) \ge \sum\limits_{k = 1}^K {{c_k}} ,\label{p1_c2}\\
&{c_{k}} \ge 0,\forall k \in {\cal K},\label{p1_c3}\\
&{\left\| {{\bf{\hat w}} \odot \left( {\widehat {\bf{d}}{{\widehat {\bf{d}}}^T}} \right)} \right\|^2} \le {P_{\rm{t}}},\label{p1_c4}\\
&{\widehat {\bf{w}}^T}{{\bf{a}}_n}{\bf{a}}_n^T\widehat {\bf{w}} \le \min \left\{ {b^2,{{\left( {{I_{\rm{H}}} - b} \right)}^2}} \right\},\forall n \in {\cal{N}}, \label{p1_c5}
\end{align}
\end{subequations}
where $t_k$  is an auxiliary variable, $\forall k \in {\cal K}$.

Then, we employ the SDR technique to relax the constraints of problem \eqref{p1}.
Constraints \eqref{p1_c1} and \eqref{p1_c2} can be   respectively  rewritten as follows
\begin{subequations}
\begin{align}
&\underbrace {\frac{1}{2}{{\log }_2}\left( {2\pi \sigma _k^2 + {\rm{Tr(}}\widehat {\bf{W}}{{\widehat {\bf{G}}}_{{\rm{c}},k}}{\rm{)}}} \right)}_{{\rm{concave}}}\nonumber\\
& - \underbrace {\frac{1}{2}{{\log }_2}\left( {2\pi \sigma _k^2 + {\rm{Tr(}}\widehat {\bf{W}}{{\widehat {\bf{G}}}_{{\rm{p}},k}}{\rm{)}}} \right)}_{{\rm{concave}}} \ge {t_k} - {c_k},\label{DC_a}\\
&\underbrace {\frac{1}{2}{{\log }_2}\left( {2\pi \sigma _k^2 + {\rm{Tr(}}\widehat {\bf{W}}{{\bf{G}}_{{\rm{c}},k}}{\rm{)}}} \right)}_{{\rm{concave}}}\nonumber\\
&- \underbrace {\frac{1}{2}{{\log }_2}\left( {2\pi \sigma _k^2 + {\rm{Tr(}}\widehat {\bf{W}}{{\overline {\bf{G}} }_{{\rm{c}},k}}{\rm{)}}} \right)}_{{\rm{concave}}} \ge \sum\nolimits_{k = 1}^K {{c_k}} .\label{DC_b}%
\end{align}
\end{subequations}

  Then,  by using the first-order Taylor
series expansion, we have
\begin{subequations}\label{DC_form1}
\begin{align}
&\frac{1}{2}{\log _2}\left( {2\pi \sigma _k^2 + {\rm{Tr(}}\widehat {\bf{W}}{{\widehat {\bf{G}}}_{{\rm{c}},k}}{\rm{)}}} \right) - {L_{{\rm{p}},k}}\left( {\widehat {\bf{W}}} \right) \ge {t_k} - {c_k},\label{DC_form1_a}\\
&\frac{1}{2}{\log _2}\left( {2\pi \sigma _k^2 + {\rm{Tr(}}\widehat {\bf{W}}{{\bf{G}}_{{\rm{c}},k}}{\rm{)}}} \right) - {L_{{\rm{c}},k}}\left( {\widehat {\bf{W}}} \right) \ge \sum\limits_{k = 1}^K {{c_k}} ,\label{DC_form1_b}
\end{align}
\end{subequations}
where ${L_{{\rm{p}},k}}$ and ${L_{{\rm{c}},k}}$ are  linear functions of variable ${\widehat {\bf{W}}}$, which are given by
\begin{subequations}
\begin{align}
{L_{{\rm{p}},k}}\left( {\widehat {\bf{W}}} \right) =& \frac{1}{2}{\log _2}\left( {2\pi \sigma _k^2 + {\rm{Tr(}}{{\widehat {\bf{W}}}^{\left[ m \right]}}{{\widehat {\bf{G}}}_{{\rm{p}},k}}{\rm{)}}} \right)\nonumber\\
 &+ \frac{{{\rm{Tr}}\left( {{{\widehat {\bf{G}}}_{{\rm{p}},k}}(\widehat {\bf{W}} - {{\widehat {\bf{W}}}^{\left[ m \right]}})} \right)}}{{\left( {2\pi \sigma _k^2 + {\rm{Tr(}}{{\widehat {\bf{W}}}^{\left[ m \right]}}{{\widehat {\bf{G}}}_{{\rm{p}},k}}{\rm{)}}} \right)2\ln 2}},\\
{L_{{\rm{c}},k}}\left( {\widehat {\bf{W}}} \right) =& \frac{1}{2}{\log _2}\left( {2\pi \sigma _k^2 + {\rm{Tr(}}{{\widehat {\bf{W}}}^{\left[ m \right]}}{{\overline {\bf{G}} }_{{\rm{c}},k}}{\rm{)}}} \right)\nonumber\\
 &+ \frac{{{\rm{Tr}}\left( {{{\overline {\bf{G}} }_{{\rm{c}},k}}(\widehat {\bf{W}} - {{\widehat {\bf{W}}}^{\left[ m \right]}})} \right)}}{{\left( {2\pi \sigma _k^2 + {\rm{Tr(}}{{\widehat {\bf{W}}}^{\left[ m \right]}}{{\overline {\bf{G}} }_{{\rm{c}},k}}{\rm{)}}} \right)2\ln 2}},
\end{align}
\end{subequations}
where ${\widehat {\bf{W}}^{\left[ m \right]}}$ is a feasible point obtained from the $m$th iteration.

Therefore, at the $\left( {m + 1} \right)$th  iteration, the  convex approximation form of  problem \eqref{p1} is given  as
\begin{subequations}\label{p2}
\begin{align}
\mathop {\max }\limits_{\scriptstyle\left\{ {{t_k},{c_k}} \right\}_{k = 1}^K\hfill\atop
\scriptstyle\;\;\;\widehat {\bf{W}}\hfill} &\sum\limits_{k = 1}^K {{t_k}}   \\
\rm{s.t.} & \eqref{DC_form1_a},~\eqref{DC_form1_b}\nonumber\\
&{\rm{Tr}}\left( {\widehat {\bf{W}} \odot (\widehat {\bf{d}}{{\widehat {\bf{d}}}^T})} \right) \le {P_{\rm{t}}},\\
&{\rm{Tr}}\left( {\widehat {\bf{W}}{{\bf{a}}_n}{\bf{a}}_n^T} \right) \le \min \left\{ {b^2,{{\left( {{I_{\rm{H}}} - b} \right)}^2}} \right\},\forall n \in {\cal{N}},\\
&\widehat {\bf{W}}\succeq {\bf{0}},{{c}_k} \ge 0,\forall k \in {\cal{K}}.
\end{align}
\end{subequations}

\section{SIMULATION RESULTS AND DISCUSSION}
Consider a RSMA-aided multi-user VLC network deployed in a room with a size of $7 \times 7 \times 5{{\rm{m}}^3}$, where the unit of distances is meter.

\begin{figure}[h]
    \begin{minipage}[b]{0.45\textwidth}
      \centering
	\includegraphics[width=8cm]{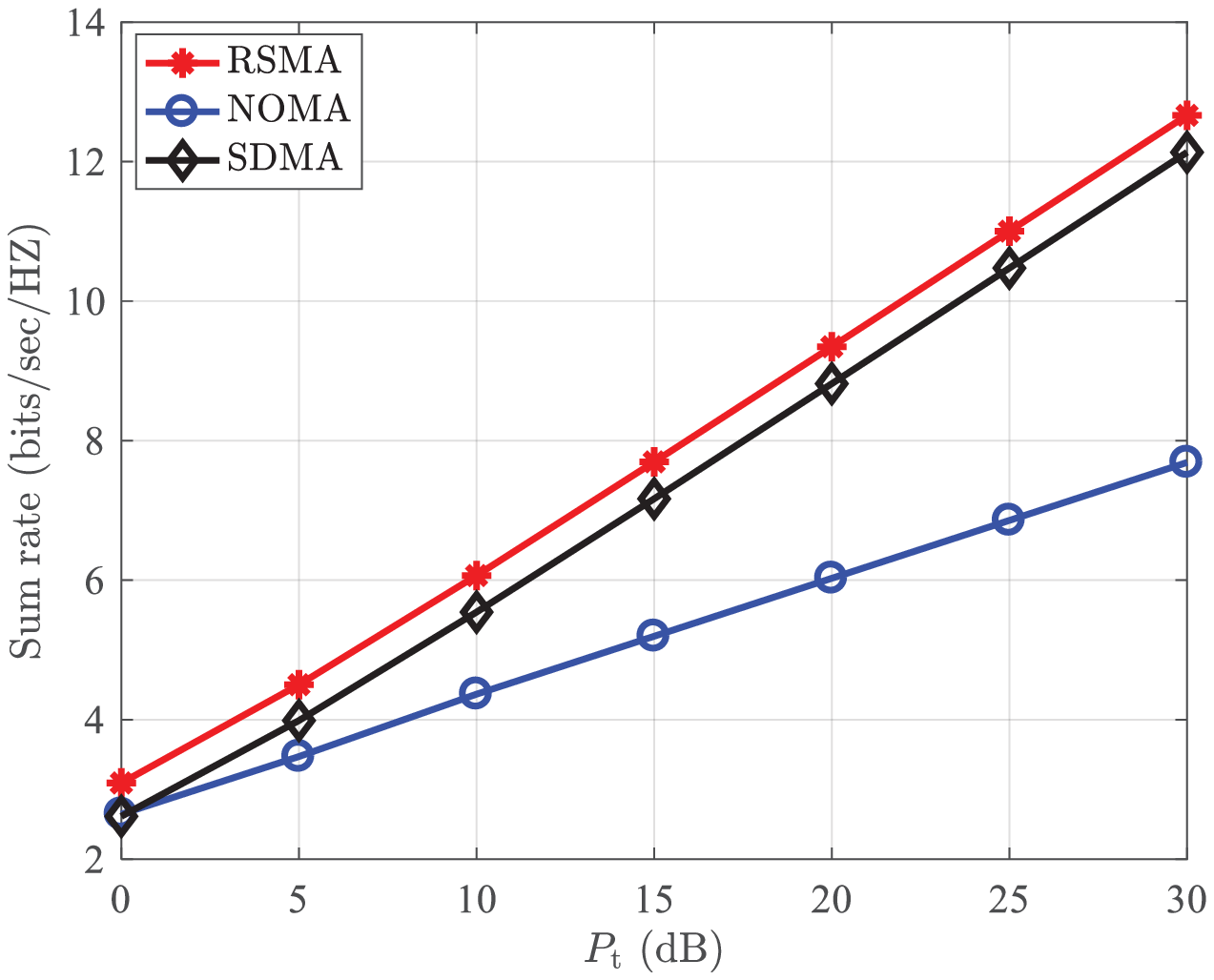}
      \vskip-0.2cm\centering {\footnotesize (a)}
    \end{minipage}\hfill
    \begin{minipage}[b]{0.45\textwidth}
    \centering
	\includegraphics[width=8cm]{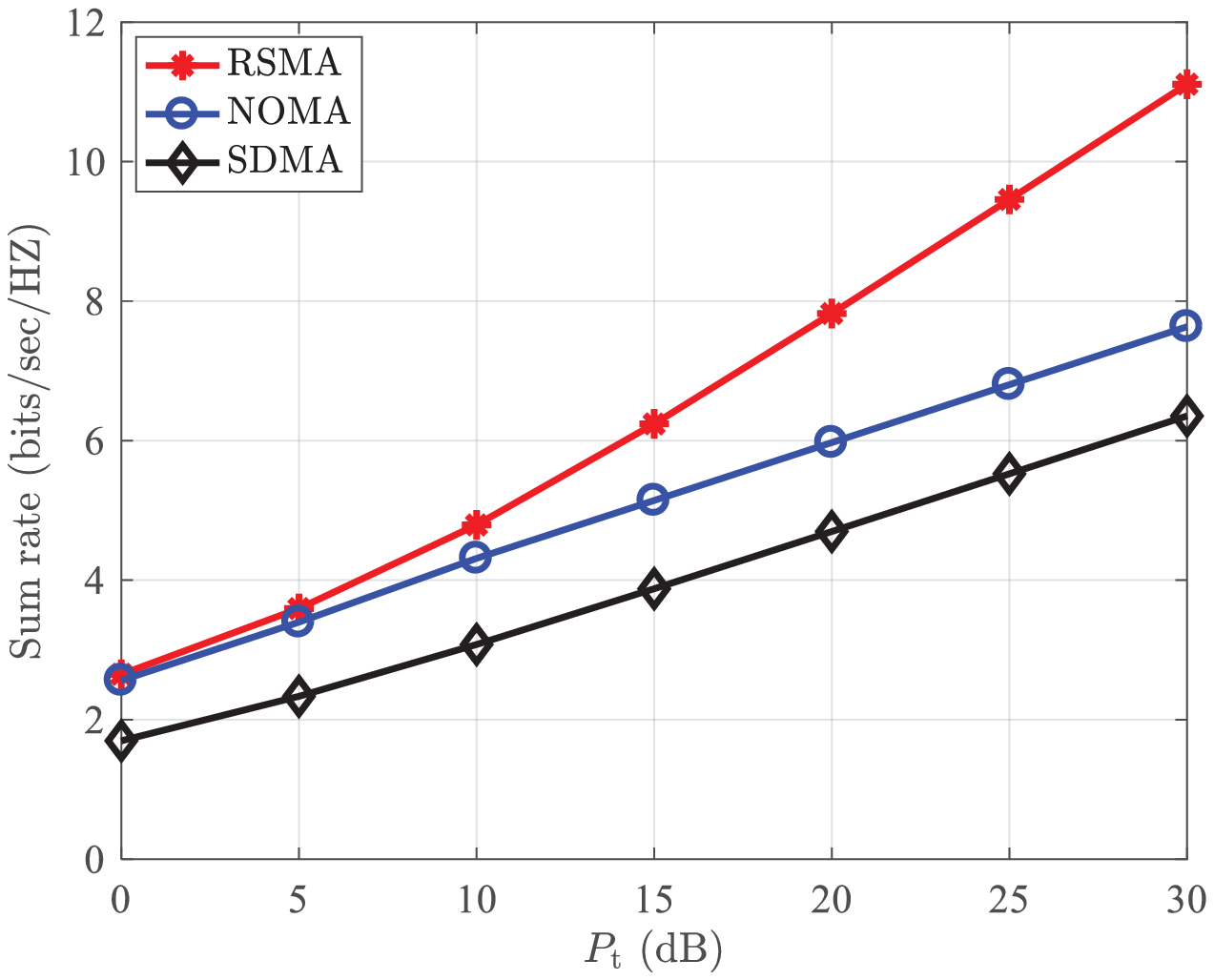}
      \vskip-0.2cm\centering {\footnotesize (b)}
    \end{minipage}\hfill
 \caption{(a)~ The sum rate (bits/sec/Hz) versus ${P_{\rm{t}}}({\rm{dB}})$ with perfect CSIT , where $N=4$, $K=2$ uses are located at $U_1$ and $U_2$;
  (b)~The sum rate $\left( {{\rm{bit/sec/Hz}}} \right)$ versus ${P_{\rm{t}}}$ $\left( {{\rm{dB}}} \right)$ with perfect CSIT, where $N=4$, $K=2$ uses are located at $U_3$ and $U_4$;}
  \label{underload} 
\end{figure}

 Fig. \ref{underload} (a)  shows the sum rate (bits/sec/Hz) of  $U_1$ and $U_2$  versus the transmitted power budget $P_{\rm{t}}$ (dB) of the RSMA, NOMA and SDMA schemes.  It can be observed that the sum rate of the three schemes increases as  $P_{\rm{t}}$ increases, and the sum rate of RSMA is higher than that of NOMA and SDMA schemes.
Furthermore, Fig. \ref{underload} (b) depicts the   sum rate of  two users $U_3$ and $U_4$   for different transmitted power budget $P_{\rm{t}}$  of the RSMA, NOMA and SDMA schemes.
  From Fig. \ref{underload} (b), we observe that  the RSMA scheme always outperforms both the NOMA   and SDMA schemes, which is similar to that in  Fig. \ref{underload} (a).
%

\section{Conclusion}

In this work, we addressed optimal and robust beamformer  design allocation
for RSMA-aided VLC networks.
Specifically, we derived the  lower bound of achievable rates of RSMA-aided VLC networks.
Moreover, based on the derived bound of achievable rates, we investigated the optimal beamformer design to maximize the sum rate under the optical and electrical power constraints of LEDs.    Simulation results illustrated the superiority of the proposed beamformer design and provide useful insights on the design of RSMA-aided VLC networks.

\bibliographystyle{IEEE-unsorted}
\bibliography{refs0611}

\end{document}